# Interpretations of Negative Probabilities

M. Burgin

Department of Mathematics
*University of California, Los Angeles*
405 Hilgard Ave.
Los Angeles, CA 90095

## Abstract

In this paper, we give a frequency interpretation of negative probability, as well as for extended probability, demonstrating that to a great extent these new types of probabilities, behave as conventional probabilities. Extended probability comprises both conventional probability and negative probability. The frequency interpretation of negative probabilities gives supportive evidence to the axiomatic system built in (Burgin, 2009) for extended probability as it is demonstrated in this paper that frequency probabilities satisfy all axioms of extended probability.

## 1. Introduction

Probability theory is nowadays an important tool in physics and information theory, engineering and industry. A great discovery of twentieth century physics was the probabilistic nature of physical phenomena at microscopic scales, described in quantum mechanics and quantum field theory. At present there is a firm consensus among the physicists that probability theory is necessary to describe quantum phenomena.

At the same time, problems of physics brought physicists to the necessity to use not only classical probability but also negative probability. Negative probabilities emerged in physics in 1930s when Dirac (1930) and Heisenberg (1931) introduced probability distributions with negative values within the context of quantum theory. However, both physicists missed its significance and possibility to take negative values, using this distribution as an approximation to the full quantum description of a system such as the atom. Wigner (1932) came to the conclusion that quantum corrections often lead to negative probabilities while he was supplanting the wavefunction from Schrödinger's equation with a probability distribution in phase space. To do this, he introduced a function, which looked like a conventional probability distribution and has later been better known as the Wigner quasi-probability distribution because in contrast to conventional probability distributions, it took negative values, which could not be eliminated or made nonnegative. Dirac (1942) not only supported Wigner's approach but also introduced the physical concept of negative energy. He wrote:

*"Negative energies and probabilities should not be considered as nonsense. They are well-defined concepts mathematically, like a negative of money."*

Richard Feynman in his keynote talk on Simulating Physics with Computers said:

*"The only difference between a probabilistic classical world and the equations of the quantum world is that somehow or other it appears as if the probabilities would have to go negative … "*

Feyman (1987) studied negative probability and discussed differentexamples demonstrating how negative probabilities naturally exist in physics and beyond.

Based on mathematical ideas from classical statistics and modern ideas from information theory, Lowe (2004/2007) also argues that the use of non-positive probabilities is both inevitable and natural.

After this, negative probabilities a little by little have become a popular although questionable technique in physics. Many physicists have used negative probabilities to solve physical problems (cf., for example, (Dirac, 1930; 1942; Heisenberg, 1931; Wigner, 1932; Sokolovski and Connor, 1991; Youssef, 1994; 1995; 2001; Scully, et al, 1994; Khrennikov, 1995; 1997; Han, et al, 1996; Curtright and Zachos, 2001; Sokolovski, 2007; Bednorz and Belzig, 2009; Hofmann, 2009)).

It is necessary to remark that are also used in machine learning Lowe, D. (2004/2007) and mathematical finance (Haug, 2007).

Mathematical problems with negative probabilities were also studied. Bartlett (1945) worked out the mathematical and logical consistency of negative probabilities. However, he did not establish rigorous foundation for negative probability utilization. Khrennikov (2009) developed mathematical theory of negative probabilities in the framework of *p*-adic analysis. This is adequate not for the conventional physics in which the majority of physicists work but only for the so-called *p*-adic physics.

The mathematical grounding for the negative probability in the real number domain was developed by Burgin (2009) who constructed a Kolmogorov type axiom system building a mathematical theory of extended probability as a probability function, which is defined for random events and can take both positive and negative real values. As a result, extended probabilities include negative probabilities. It was also demonstrated that the classical probability is a positive section (fragment) of extended probability.

At the same time, there are many problems with interpretations of the conventional concept of probability. The goal of this paper is to build the frequency interpretation of extended probability because the frequency interpretation reflects the most popular approach to probability treatment in science, in particular, in physics, and in other applications of probability. In Section 2, going after Introduction, traditional approaches to interpretations of probability are discussed. In Section 3, the frequency interpretation of extended probability is constructed. It is called the frequency interpretation probability. In Section 4, it is proved that the frequency interpretation probability satisfies all axioms from (Burgin, 2009).

## 2. Traditional approaches to interpretations of probability

An interpretation of the concept of probability is a choice of some class of events (or statements) and an assignment of some meaning to probability claims about those events (or statements). Usually researchers concentrate on three main interpretations of the probability: the *frequency interpretation*, the *belief interpretation*, and the *support interpretation*. There are also other interpretations, such as the logical interpretation or the propensity interpretation.

All these interpretations belong to three groups:
- *Objective probability* is defined (exists) as a numerical property of sequences of frequencies associated with an event in natural, social or technical phenomena.
- *Subjective probability* is defined (exists) as a belief or measure of confidence in an outcome of a certain event.
- *Combined probability* is defined (exists) as a belief supported by both observational and experimental evidence or measure of confidence in an outcome of some event.

**Example 2.1**. When there are 10 blue balls, 5 green balls and 5 red balls in a box, and these balls are well mixed, then the objective probability to draw a blue ball from this box is ½ .

**Example 2.2**. When we believe that the ratio of blue balls to all balls in a box is ½ , then the subjective probability of drawing a blue ball from this box is ½ .

**Example 2.3**. When we conjecture that the ratio of blue balls to all balls in a box is ½ , we tested this hypothesis and got some experimental evidence in support of it, then the combined probability of drawing a blue ball from this box is ½ .

Each of these three groups can be divided into several subclasses.

It is possible to interpret objective probability in three main ways: as the actual finite relative frequency (the *actual frequency interpretation*), in a form of a limit (or limiting behavior) of hypothetical infinite relative frequencies (the *potential frequency interpretation*) or as a propensity (the *propensity interpretation*).

It is possible to interpret subjective probability in three main ways: as an actual belief, or system of actual beliefs, (the *belief interpretation*), as an idealized belief, system of idealized beliefs, based on the Bayes theorem (the *Bayesian* or *personalist interpretation*) or in a form of a logical system (the *logical interpretation*).

It is possible to interpret combined probability in three main ways: as an actual belief, or system of actual beliefs, supported by observational or experimental evidence (the *supported belief interpretation*), as an idealized belief, system of idealized beliefs, based on the Bayes theorem and supported by observational or experimental evidence (the *supported Bayesian interpretation*) or in a form of a logical system supported by observational or experimental evidence (the *supported logical interpretation*).

It is important to distinguish two types of probabilities: *ensemble probabilities* and *sequential* (or *cumulative*) *probabilities*. A special case of sequential (or cumulative) probabilities is *temporal probability* when the sequence of events is consequential, i.e., events happen one after another.

In the *actual frequency interpretation*, the probability $p(A)$ of an event $A$ is taken equal to the long-run relative frequency with which A occurs in identical repeats of an experiment or observation.

In the *potential frequency interpretation*, the probability $p(A)$ of an event $A$ is taken equal to the limit of relative frequency with which A occurs in identical repeats of an experiment or observation.

A *long-run propensity theory* is one in which propensities are associated with repeatable conditions, and are regarded as propensities to produce in a long series of repetitions of these conditions. In this theory, frequencies are approximately equal to the probabilities. A single-case propensity theory is one in which propensities are regarded as propensities to produce a particular result on a specific occasion.

In the *Bayesian interpretation*, the probability $p(A/B)$ of an event (proposition/hypothesis) *A*, given (conditional on) the happening (truth of) the event (proposition/hypothesis) *B* is a measure of the plausibility of the event (proposition/hypothesis) *A*, given (conditional on) the happening (truth of) the event (proposition/hypothesis) *B*. Bayesian inference uses probability distribution as an encoding of our uncertainty about some model parameter or set of competing theories, based on our current state of information.

The supported Bayesian approach in the style of de Finetti (1937) recognizes no rational constraints on subjective probabilities beyond:

1. conformity to the probability calculus (coherence);
2. a rule for updating probabilities in the face of new evidence (*conditioning*).

Conditioning means that an agent with probability function $P_1$, who becomes certain of a piece of evidence *E*, should shift to a new probability function $P_2$ related to $P_1$ by:

(Conditioning) $P_2(X) = P_1(X \mid E)$ (provided $P_1(E) > 0$).

Frequency approach was mathematically grounded and further developed in algorithmic information theory (Kolmogorov, 1965; Martin-Löf, 1970).

Considering different applications of probability theory, we come to the fundamental question: what interpretation of probability is appropriate for scientific practice? One of the most outstanding philosophers of science, Rudolf Carnap suggested that both the objective (frequency) and subjective (for Carnap (1950), logical) interpretations are needed for representing different aspects of probability usage.

However, reality shows that there is no easy compromise. Indeed, most scientists (and some philosophers) support the frequency approach and do not trust the Bayesian approach oriented to subjective evaluations. At the same time, many philosophers (and some scientists) are subjectivists, supporting Bayesian approach and many of them do not believe in objective probabilities.

All approaches have their caveats. The frequency interpretation contains the term ''identical repeats.'' Of course the repeated experiments can never be identical in all respects. The Bayesian definition of probability involves the rather vague sounding term ''plausibility,'' which must be given a precise meaning for the theory to provide quantitative results.

Problems with probability interpretations and necessity to have sound mathematical foundations brought forth an axiomatic approach in probability theory. Based on ideas of Fréchet and following the axiomatic mainstream in mathematics, Kolmogorov developed his famous axiomatic exposition of probability theory (1933).

Here we consider only the frequency interpretation as the most popular approach to probability in physics because negative probability comes from physics and is more and more use by physicists.

### 3. Frequency interpretations of extended probability

Extended probabilities generalize the standard definition of a probability function, allowing operation with negative probabilities. At first, we define extended probabilities as limits of relative frequencies and then we show that so defined extended probabilities satisfy all axioms from (Burgin, 2009).

To define extended probability, we need some concepts and constructions, which are described below.

Let us consider a set $\Omega$, which consists of two irreducible parts (subsets) $\Omega^+$ and $\Omega^-$, i.e., neither of these parts is equal to its proper subset, a set **F** of subsets of $\Omega$, and a function $P$ from **F** to the set **R** of real numbers.

Let us define the union with annihilation of two subsets $X$ and $Y$ of $\Omega$ by the following formula:

$$X + Y \equiv (X \cup Y) \setminus [(X \cap -Y) \cup (-X \cap Y)]$$

Here the set-theoretical operation \ represents annihilation, while sets $X \cap -Y$ and $X \cap -Y$ depict annihilating entities.

Note that annihilation occurs not only in physics where particles and antiparticles annihilate one another, but also in ordinary life of people. For instance, a person has stocks of two companies. If in 2009, the first set of stocks gave profit $1,000, while the second set of stocks dropped by $1,000, then the combined income was $0. The loss annihilated the profit.

Here is even a simpler example. A person finds $20 and looses $10. As the result, the amount of this person's money has increased by $10. The loss annihilated part of the increase.

Thus, it is natural to assume that if $A$ belongs to $\mathbf{F}$, then it does contain pairs $\{w, -w\}$ with $w \in \Omega$ and $\mathbf{F}^+ = \{X \in \mathbf{F}; X \subseteq \Omega^+\}$ is a set algebra (cf., for example, (Kolmogorov and Fomin, 1989)) with respect to union with annihilation and $\Omega^+$ is a member of $\mathbf{F}^+$.

Elements from $\mathbf{F}$, i.e., subsets of $\Omega$ that belong to $\mathbf{F}$, are called *random events*.

Elements from $\mathbf{F}^+ = \{X \in \mathbf{F}; X \subseteq \Omega^+\}$ are called *positive random events*.

Elements from $\Omega^+$ that belong to $\mathbf{F}^+$ are called *elementary positive random events* or simply, *elementary positive random events*.

If $w \in \Omega^+$, then $-w$ is called the *antievent* of $w$. We assume that $-(-w) = w$.

Elements from $\Omega^-$ that belong to $\mathbf{F}^-$ are called *elementary negative random events* or *elementary random antievents*.

For any set $X \subseteq \Omega$, we define

$$X^+ = X \cap \Omega^+,$$
$$X^- = X \cap \Omega^-,$$
$$-X = \{-w; w \in X\}$$

and

$$\mathbf{F}^- = \{-A; A \in \mathbf{F}^+\}$$

If $A \in \mathbf{F}^+$, then $-A$ is called the *antievent* of $A$.

We assume that $\mathbf{F} \equiv \{X; X^+ \subseteq \mathbf{F}^+ \ \& \ X^- \subseteq \mathbf{F}^- \ \& \ X^+ \cap -X^- \equiv \emptyset \ \& \ X^- \cap -X^+ \equiv \emptyset\}$.

Elements from $\mathbf{F}^-$ are called *negative random events* or *random antievents*.

Here we treat only the finite case when $\Omega = \{w_1, w_2, w_3, \ldots, w_n, -w_1, -w_2, -w_3, \ldots, -w_n\}$, $\Omega^+ = \{w_1, w_2, w_3, \ldots, w_n\}$ and $\Omega^- = \{-w_1, -w_2, -w_3, \ldots, -w_n\}$.

Taking an event $\{u_i\}$ with $u_i \in \Omega$, we denote by $N_+$ the number times that events with the same sign as the event $u_i$ occur during a sequence of $N$ trials, by $N_-$ the number times that events with the opposite to the event $u_i$ sign occur during a sequence of $N$ trials, by $n_i$ the number of times that the event $u_i$ occurs during a sequence of $N$ trials, and by $m_i$ the number of times that the event $-u_i$ occurs during the same sequence of trials. Let $v_N(u_i) = (n_i)/N_+ - (m_i)/N_-$. Then we define the extended frequency probability of the event $u_i$ as

$$p(u_i) = \lim_{N \to \infty} v_N(u_i)$$

We assume that this limit exists for random events.

In this formula $u_i$ is equal to $w_i$ or to $-w_i$. Thus, taking an event $\{w_i\}$ with $w_i \in \Omega^+$, we have the number $N_+$ of times that positive events occur during a sequence of $N$ trials, the number $N_-$ of times that negative events occur during a sequence of $N$ trials, the number $n_i$ of times that the event $w_i$ occurs during a sequence of $N$ trials and the number $m_i$ of times that the event $-w_i$ occurs during the same sequence of trials. Then we have

$$v_N(w_i) = (n_i)/N_+ - (m_i)/N_-$$

and the extended frequency probability of the event $w_i$ is equal to

$$p(w_i) = \lim_{N \to \infty} v_N(w_i)$$

For an event $\{-w_i\}$ with $-w_i \in \Omega^-$, we have the number $m_i$ of times that the event $-w_i$ occurs during a sequence of $N$ trials and the number $n_i$ of times that the event $-w_i$ occurs during the same sequence of trials. Then $v_N(-w_i) = (m_i)/N_- - (n_i)/N_+$ and the extended frequency probability of the event $w_i$ is equal to

$$p(-w_i) = \lim_{N \to \infty} v_N(-w_i)$$

Consequently, we have

$$p(w_i) = -p(-w_i)$$

In a general case of random events, we have a random event $A = \{w_{i1}, w_{i2}, w_{i3}, \ldots, w_{ik}, -w_{j1}, -w_{j2}, -w_{j3}, \ldots, -w_{jt}\}$, the number $N_+$ of times that positive events occur during a sequence of $N$ trials, the number $N_-$ of times that negative events occur during a sequence of $N$ trials, the number $n_{pos}$ of times that positive events $w_{ir}$ from $A$ occur during a sequence of $N$ trials and the number $n_{neg}$ of times that negative events $-w_{iq}$ occur during the same sequence of trials.

Then we define $v_N(A) = (n_i)/N_+ - (m_i)/N_-$ and the extended frequency probability of the random event $A$ as

$$p(A) = \lim_{N \to \infty} v_N(A)$$

We assume that this limit exists for random events, i.e., for all events from the set **F**.

In other words, when the number $N$ of trials goes to infinity ($N \to \infty$) the number $v_N(A)$ approaches the *probability* of event $A$. The regularity of $v_N(A)$ converging to a proper fraction characterizes the meaning of the probability of event $A$. The words *same random experiment* imply that an identical random experiment (in every conceivable respect) is duplicated infinitely often under unchanging circumstances.

This interpretation is similar to the traditional relative frequency interpretation of probability and in the case when there are no negative elementary events, we exactly obtain the traditional relative frequency interpretation of probability, where the ratio $n_i/N$ approaches the *probability* of event $A$ when the number $N$ of trials goes to infinity.

To better understand how negative elementary events appear and how negative probability emerges, consider the following example.

Let us consider the situation when an attentive person $A$ with the high knowledge of English writes some text $T$. We may ask what the probability is for the word "texxt" or "wrod" to appear in his text $T$. Conventional probability theory gives 0 as the answer. However, we all know that there are usually misprints. So, due to such a misprint this word may appear but then it would be corrected. In terms of extended probability, a negative value (say, -0.1) of the probability for the word "texxt" to appear in his text $T$ means that this word may appear due to a misprint but then it'll be corrected and will not be present in the text $T$.

Negative probability becomes even less (say, -0.3) when people use word processors because misprints become more probable. For instance, it is possible to push a wrong key on the keyboard or pushing one key also to push its neighbor.

In physics, negative probability may reflect the situation when instead of a particle its anti-particle appears. For instance, probability -0.3 that in a given interaction an electron appears means that there is probability 0.3 that in this interaction a positron appears.

Note that if we have an event $A = \{ w, -w \}$ with $\{w\} \in \mathbf{F}$, its extended frequency probability is equal to zero although by conditions on $\mathbf{F}$, the event $A$ does not belong to $\mathbf{F}$. Indeed,

$$p(A) = \lim_{N \to \infty} v_N(A)$$

At the same time, as $-(-w) = w$, we have

$$v_N(A) = v_N(w) + v_N(-w) =$$

$$((n)/N_+ - (m)/N_-) + ((m)/N_+ - (n)/N_-) = 0$$

where $n$ is the number of times that the event $w$ occurs during a sequence of $N$ trials and $m$ is the number of times that the event $-w$ occurs during the same sequence of trials. Thus,

$$p(A) = 0$$

Similar reasoning shows that if some event $B$ contains elements $w$ and $-w$, then this pair does not influence the extended frequency probability $p(B)$ of the event $B$ and both elements $w$ and $-w$ may be annihilated.

**4. Axioms for extended probability and their validity for frequency probability**

Let us consider axioms for extended probability from (Burgin, 2009).

**Definition 1.** The function $P$ from $\mathbf{F}$ to the set $\mathbf{R}$ of real numbers is called a *probability function*, if it satisfies the following axioms:

**EP 1** (*Order structure*).  There is a graded involution $\alpha: \Omega \to \Omega$, i.e., a mapping such that $\alpha^2$ is an identity mapping on $\Omega$ with the following properties: $\alpha(w) = -w$ for any element $w$ from $\Omega$, $\alpha(\Omega^+) \supseteq \Omega^-$, and if $w \in \Omega^+$, then $\alpha(w) \notin \Omega^+$.

**EP 2** (*Algebraic structure*). $\mathbf{F}^+ \equiv \{X \in \mathbf{F}; X \subseteq \Omega^+\}$ is a set algebra that has $\Omega^+$ as a member.

**EP 3** (*Normalization*). $P(\Omega^+) = 1$.

**EP 4** (*Composition*) $\mathbf{F} \equiv \{X; X^+ \subseteq \mathbf{F}^+ \ \& \ X^- \subseteq \mathbf{F}^- \ \& \ X^+ \cap -X^- \equiv \emptyset \ \& \ X^- \cap -X^+ \equiv \emptyset\}$.

**EP 5** (*Finite additivity*)
$$P(A \cup B) = P(A) + P(B)$$

for all sets $A, B \in \mathbf{F}$ such that

$$A \cap B \equiv \emptyset$$

As it is proved in (Burgin, 2009) that $P(\Omega^+) = 1$, axioms EP 3 and EP 5 imply that extended probability takes values in the interval $[-1, 1]$.

**EP 6** (*Annihilation*). $\{v_i, w, -w; v_i, w \in \Omega \ \& \ i \in I\} = \{v_i; v_i \in \Omega \ \& \ i \in I\}$ for any element $w$ from $\Omega$.

Axiom EP6 shows that if $w$ and $-w$ are taken (come) into one set, they annihilate one another. Having this in mind, we use two equality symbols: $=$ and $\equiv$. The second symbol means equality of elements of sets. The second symbol also means equality of sets, when two sets are equal when they have exactly the same elements (Kuratowski and Mostowski, 1967). The equality symbol $=$ is used to denote equality of two sets with annihilation, for example, $\{w, -w\} = \emptyset$. Note that for sets, equality $\equiv$ implies equality $=$.

For equality of numbers, we, as it is customary, use symbol $=$.

**EP 7**. (*Adequacy*)  $A = B$ implies $P(A) = P(B)$ for all sets $A, B \in \mathbf{F}$.

For instance, $P(\{w, -w\}) = P(\emptyset) = 0$.

**EP 8**. (*Non-negativity*) $P(A) \geq 0$, for all $A \in \mathbf{F}^+$.

As in the case of the classical probability, it is possible to add one more axiom called *axiom of continuity* to the list of axioms for extended probability. This allows us to comply with the traditional approach to probability.

**EP 9**. (*Continuity*) If

$$A_1 \supseteq A_2 \supseteq A_3 \supseteq \ldots \supseteq A_i \supseteq \ldots$$

is a decreasing sequence of events $A_i$ from $\mathbf{F}^+$ such that

$$\bigcap_{i=1}^{\infty} A_i = \emptyset,$$

then

$$\lim_{i \to \infty} P(A_i) = 0.$$

It is possible to consider a restricted form of Axiom EP5.

**EP 10** (*Decomposition*) For any $A \in \mathbf{F}$, we have
$$P(A) = P(A^+) + P(A^-)$$

Let us check axioms EP 1 – 8, 10 for extended frequency probability. We do not consider Axiom EP 9 because here we treat only the case with a finite set $\Omega$ of elementary events and in the finite case, Axiom EP 9 follows from the equality $P(\emptyset) = 0$.

**Theorem 1**. Extended frequency probability $p$ satisfies Axioms EP 1 – 8, 10.

Proof. **Axiom EP 1**. The graded involution $\alpha: \Omega \to \Omega$ is defined by the rule $\alpha(w_i) = -w_i$ for all $i = 1, 2, 3, \ldots, n$. Then $\alpha(\Omega^+) = \Omega^-$, and if $w \in \Omega^+$, then $\alpha(w) \notin \Omega^+$.

**Axiom EP 2** is true because $\mathbf{F}$ is defined so that $\mathbf{F}^+$ is a set algebra with respect to union with annihilation and has $\Omega^+$ as a member.

**Axiom EP 3.** By definitions, we have

$$v_N(\Omega^+) = (n_i)/N_+ - (m_i)/N_-$$

As any positive elementary event belongs to $\Omega^+$, we have $n_i = N_+$. Thus, $(n_i)/N_+ = 1$.

Besides, $m_i = 0$ as does not have negative elementary events. Thus, $(m_i)/N_- = 0$.

Consequently,

$$v_N(\Omega^+) = 1 \text{ for all } N$$

and

$$p(\Omega^+) = \lim_{N \to \infty} v_N(\Omega^+) = 1$$

**Axiom EP 4** is true by assumptions on **F**.

**Axiom EP 5.** By definition, we have

$$p(A) = \lim_{N \to \infty} v_N(A) =$$

$$\lim_{N \to \infty} ((n_{pos})/ N_+) - \lim_{N \to \infty}(( n_{neg})/ N_+)$$

and

$$p(B) = \lim_{N \to \infty} v_N(B) =$$

$$\lim_{N \to \infty} ((r_{pos})/ N_+) - \lim_{N \to \infty}(( r_{neg})/ N_+)$$

where $N_+$ is the number of times that positive events occur during a sequence of $N$ trials, $N_-$ is the number of times that negative events occur during a sequence of $N$ trials, $n_{pos}$ is the number of times that positive events from $A$ occur during a sequence of $N$ trials, $n_{neg}$ is the number of times that negative events from $A$ occur during a sequence of $N$ trials, $r_{pos}$ is the number of times that positive events from $B$ occur during a sequence of $N$ trials, and $r_{neg}$ is the number of times that positive events from $B$ occur during a sequence of $N$ trials,. Thus,

$$p(A \cup B) = \lim_{N \to \infty} v_N(A \cup B) =$$

$$\lim_{N \to \infty} ((((n_{pos})/ N_+) + (r_{pos})/ N_+)) - (((n_{neg})/ N_-) + ((r_{neg})/ N_-)))$$

because $A$ and $B$ do not have common elements. Consequently, we have

$$\lim_{N \to \infty} ((((n_{pos})/ N_+) + (r_{pos})/ N_+)) - (((n_{neg})/ N_-) + ((r_{neg})/ N_-))) =$$

$$\lim_{N \to \infty} (((n_{pos})/ N_+) - ((n_{neg})/ N_-) + (r_{pos})/ N_+)) - (r_{neg})/ N_-))=$$

$$\lim_{N \to \infty} ((((n_{pos})/ N_+) - ((n_{neg})/ N_-)) + \lim_{N \to \infty} (((r_{pos})/ N_+)) - (r_{neg})/ N_-))$$

$$\lim_{N \to \infty} v_N(A) + \lim_{N \to \infty} v_N(B) = p(A) + p(B)$$

Thus,

$$p(A \cup B) = p(A) + p(B)$$

**Axiom EP 6** and **EP 7** are true because the set **F** contains only subsets of $\Omega$ that do not contain opposite elementary events.

**Axioms EP 8.** If $A \in \mathbf{F}^+$, $A$ then does not contain negative elementary events. Consequently, $m_i = 0$ as does not have negative elementary events. Thus, $(m_i)/N_- = 0$. Thus, $v_N(A) = (n_i)/N_+ \geq 0$ and $p(A) = \lim_{N \to \infty} v_N(A) \geq 0$.

**Axiom EP 10.** Taking a random event $A$ from $\mathbf{F}$, we have
$$p(A) = p(A^+) + p(A^-)$$

because $p(A^+) = \lim_{N \to \infty} (n_i)/N_+$, $p(A^-) = \lim_{N \to \infty} (m_i)/N_-$, and the limit of the sum of two sequences is the sum of the limits of each of these sequences.

4. **Conclusion**

We have built the frequency interpretation of extended probabilities, which include negative probabilities and demonstrated that extended frequency probability satisfies all axioms for extended probability from (Burgin, 2009). This eliminates many problems that researchers have with negative probabilities, demonstrating consistency of the axioms that characterize extended probability and validating utilization of negative probability as a mathematically dependable concept.